\documentclass[twocolumn,superscriptaddress,pra,amsmath,amssymb]{revtex4-1}
\usepackage{graphicx}

\begin{document}

\newcommand{\erw}[1]{ {\cal E} \!\left[ #1 \right]\!}
\newcommand{\la}{\langle}
\newcommand{\ra}{\rangle}
\newcommand{\da}{^{\dagger}}
\newcommand{\diff}{\mathrm{d}}
\newcommand{\tr}{\mathrm{tr}}
\newcommand{\sinc}{\text{\rm sinc}}
\newcommand{\erf}{\text{\rm erf}}
\newcommand{\vp}{\boldsymbol{p}}
\newcommand{\vpi}{\boldsymbol{\pi}}
\newcommand{\vs}{\boldsymbol{s}}
\newcommand{\vR}{\boldsymbol{R}}
\newcommand{\vP}{\boldsymbol{P}}
\newcommand{\vq}{\boldsymbol{q}}
\newcommand{\vj}{\boldsymbol{j}}
\newcommand{\vk}{\boldsymbol{k}}
\newcommand{\vu}{\boldsymbol{u}}
\newcommand{\va}{\boldsymbol{a}}
\newcommand{\vv}{\boldsymbol{v}}
\newcommand{\vx}{\boldsymbol{x}}
\newcommand{\ex}{\boldsymbol{e}_x}
\newcommand{\ey}{\boldsymbol{e}_y}
\newcommand{\ez}{\boldsymbol{e}_z}
\newcommand{\oH}{\mathsf H }
\newcommand{\oU}{ \mathsf U }
\newcommand{\oW}{ \mathsf W }
\newcommand{\oS}{ \mathsf S }
\newcommand{\oP}{ \mathsf P }
\newcommand{\oL}{ \mathsf L }
\newcommand{\oM}{\mathsf m }
\newcommand{\oN}{ \mathsf N }
\newcommand{\on}{ \mathsf n }
\newcommand{\oa}{ \mathsf a }
\newcommand{\ovp}{ \text{\sf \textbf{p}}} 
\newcommand{\ovr}{ \text{\sf \textbf{x}}} 
\newcommand{\ovX}{ \mathsf {X} }
\newcommand{\R}{ \mathbb R }
\newcommand{\C}{ \mathbb C }
\newcommand{\cS}{ {\cal S} }
\newcommand{\cL}{ {\cal L} }
\newcommand{\U}{ {\cal U} }
\newcommand{\V}{ {\cal V} }
\newcommand{\T}{ {\cal T} }
\newcommand{\Order}{ {\cal O} }
\newcommand{\Vsin}{\V}
\newcommand{\density}{\varrho}
\newcommand{\ovP}{ \text{\sf \textbf{P}}} 
\newcommand{\ovR}{ \text{\sf \textbf{X}}} 

\title{Macroscopicity of Mechanical Quantum Superposition States}

\author{Stefan Nimmrichter}
\affiliation{University of Vienna, Vienna Center for Quantum Science and Technology (VCQ), Faculty of Physics, Boltzmanngasse 5, 1090 Vienna, Austria}

\author{Klaus Hornberger}
\affiliation{University of Duisburg-Essen, Faculty of Physics, Lotharstra{\ss}e 1, 47048 Duisburg, Germany}

\date{\today}

\begin{abstract}
We propose an experimentally accessible, 
objective measure for the macroscopicity of superposition states in mechanical quantum systems. Based on the observable consequences of a minimal, macrorealist extension of quantum mechanics, it allows one to quantify the degree of macroscopicity achieved in different experiments.
\end{abstract}

\maketitle

\emph{Introduction.---}
Experiments probing the quantum superposition principle at the borderline 
to classical mechanics are a driving force of modern physics.
This includes the demonstration of superposition states of counterrunning currents involving $10^{14}$ 
electrons \cite{Friedman2000_long,Hime2006}, of Bose-Einstein condensed atoms \cite{Andrews1997_short}, and complex molecules \cite{Gerlich2011_short}.

Various measures have been suggested for the size of superposition states involving macroscopically distinct properties of complex quantum systems\cite{Leggett1980,Leggett2002,Duer2002,Bjoerk2004,Korsbakken2007,Marquardt2008,Lee2011}. 
Most of them refer to specific types or representations of quantum states, 
or count the operational resources required to analyze 
them. 
While most proposals seem to be grounded in a common 
information-theoretic framework \cite{Froewis2012}, we still lack a method of attributing a definite and unbiased measure to all experimental tests of the quantum superposition principle.

The task to define a macroscopicity measure `within' quantum theory
is confounded by a fundamental problem: We are free to decompose a many particle Hilbert space into different tensor products, such that a complicated single-particle representation of a wave function may look mundane after a change of variables to collective degrees of freedom. This highlights the problems of an \emph{ad hoc} selection of distinguished observables.

In view of this, we propose to call a quantum state of a 
mechanical system the more macroscopic the better its experimental demonstration allows one to rule out even a minimal modification of quantum mechanics, which would predict a failure of the superposition principle on the macroscale. 
Turning this characterization into a definite measure requires one to specify the minimal modification. 

Fortunately, it is not necessary to worry about the details of possible nonlinear or stochastic additions to the Schr{\"o}dinger equation, which might embody the coarse-grained effects of a deeper theory, say, incorporating gravitation or a granular space-time \cite{Diosi1989,Penrose1996,Smolin2001a}, 
or might represent a fundamental stochasticity \cite{Ghirardi1990b,Milburn1991,Adler2004}. 
All that matters empirically are their observable consequences, described by the dynamics of the many-body density operator. 
We argue that basic consistency, symmetry, and scaling arguments lead to 
an explicit, parametrizable characterization of  
the impact of a minimally intrusive modification. 
The fact that no evidence of such physics `beyond the Schr\"odinger equation'
is seen in a quantum experiment rules out a certain parameter region. 
For a superposition state in a different experiment to be more macroscopic, its demonstration must exclude a larger parameter region, implying that possible modifications must be even weaker.
Diverse experiments can thus be compared without prejudice.

\emph{Minimal modification of quantum mechanics.---}
The modification must serve to `classicalize' the state evolution 
in the sense that superpositions of macroscopically distinct 
mechanical states are turned rapidly into mixtures.
The operational description of quantum theory,
based on the state operator $\rho$, its completely positive and trace-preserving time evolution, and a consistent rule of assigning probabilities to measurements \cite{Kraus1983states}, allows one to treat 
(nonrelativistic) 
quantum and classical mechanics in a common general formalism. 
It is therefore natural to account for an objective modification of the quantum time evolution in the framework of dynamical semigroups \cite{Alicki2007quantum}. 
That is, the effect of the modification can be expressed as a generator $\cL_N$ 
added to the von Neumann equation for the state of motion $\rho_N$ of an arbitrary system of $N$ particles, $\partial_t \rho_N=[\oH,\rho_N]/i\hbar+\cL_N \rho_N$.

In addition, 
we require the modification 
(i) to be invariant under Galilean transformations, avoiding a distinguished frame of reference, 
(ii) to leave the exchange symmetry of identical particles unaffected, 
(iii) to respect the `innocent bystander' condition that adding an uncorrelated system leaves the reduced state unchanged, 
and (iv) to display scale invariance with respect to the center-of-mass (c.m.) of a compound system.
We will see that these requirements essentially determine the form of a possible minimal modification.

Let us first consider an elementary particle of mass $m$
using the formalism of quantum dynamical semigroups.
A theorem by Holevo \cite{Holevo1993} states that any Galilean invariant addition to the 
von Neumann equation for the state of motion $\rho$ must have the form $\partial_t \rho=[\oH,\rho]/i\hbar+\cL_1 \rho$ with
\begin{align}
 \cL_1 \rho &= \frac{1}{\tau} \left[ \int \!\! \diff^3 s \diff^3 q \, g\left( s,q \right) \oW \left( \vs,\vq \right) \rho \oW\da \left( \vs,\vq \right) - \rho\right], \label{eqn:CME}
\end{align}
if one disregards unbounded diffusion terms, which would yield a substantially more drastic modification. The operators 
\begin{align}
 \oW \left( \vs,m \vv \right) &= \exp \left[ \frac{i}{\hbar} \left( \ovp \cdot \vs - m\vv \cdot \ovr \right) \right] \label{eqn:Weylop}
\end{align}
effect a translation $ \vs$ and a velocity boost $ \vv$ of the elementary particle,
while $g\left( s,q \right)$ is a positive, isotropic and normalized phase-space distribution,
whose standard deviations for the position and the momentum variable will be denoted by $\sigma_s$ and $\sigma_q$. The von Neumann equation
is reobtained for $\sigma_s=\sigma_q=0$.

The modification (\ref{eqn:CME}) serves its purpose of classicalizing the motion of a single particle: 
It effects a decay of the position and the momentum off-diagonal matrix elements of $\rho$.
The parameter $\tau$ provides the corresponding time scale for those
matrix elements 
which are more than the critical length scale
$\hbar/\sigma_q$ off the diagonal in position, or more than $\hbar/\sigma_s$ off in momentum, whereas smaller-scale coherences may survive much longer. 
Delocalized superposition states thus get localized in phase space
as time evolves, ultimately
rendering 
the phase-space representation of
$\rho$ indistinguishable from 
an equivalent
classical distribution.
At the same time, the modification (\ref{eqn:CME}) induces a position and momentum diffusion, implying that any bound particle gradually gains energy.  For harmonic binding potentials with frequency $\omega$ the energy increases as 
$\sigma_q^2/2m + m \omega^2 \sigma_s^2/2$ 
per unit of time $\tau$. 

\emph{Many-particle description.---}
The requirement of Galilean invariance in a general mechanical system of $N$ particles
implies that the phase-space translation operators 
must effect a net shift of the center-of-mass coordinates by $\vs$ and $\vv$.
On the other hand, the scale invariance conditions with respect to 
an innocent bystander (iii) and to the center-of-mass (iv) 
require that the equation for the $N$-particle state reduces to the single-particle form whether one traces over the other $N-1$ particles or over the relative coordinates in a compound object 
of rigidly bound contsituents;
in the latter case the single mass should be replaced by the total mass $M=\sum_n m_n$.
This is achieved by composing the $N$-particle operators as the weighted sum of the single-particle operators (\ref{eqn:Weylop}),
\begin{align}
 \oW_N \left( \vs,\vq \right) &= \sum_{n=1}^N \frac{m_n}{m_{\rm e}} \exp \left[ \frac{i}{\hbar} \left( \ovp_n \cdot \frac{m_{\rm e}}{m_n}\vs - \vq \cdot \ovr_n \right) \right],
 \label{eqn:WN}
\end{align}
where $m_{\rm e}$ is an arbitrary reference mass;
see the Appendix for details.

We note that the operators (\ref{eqn:WN}) conserve the exchange symmetry of a quantum state. The corresponding $N$-particle equation
\begin{align}
 \cL_N \rho_N &= \frac{1}{\tau_{\rm e}} \int \!\! \diff^3 s \diff^3 q\, g_{\rm e}\left( s,q \right) 
 \Big[ \oW_N \left( \vs,\vq \right) \rho_N \oW_N\da \left( \vs,\vq \right)  \nonumber \\
&\quad 
- \frac{1}{2} \left\{ \oW_N\da \left( \vs,\vq \right)\oW_N \left( \vs,\vq \right),\rho_N \right\} \Big], 
\label{eqn:CME_N}
\end{align}
thus leaves boson and fermion statistics invariant (ii).
The equation is completely determined once we specify the mass $m_{\rm e}$, the coherence time parameter $\tau_{\rm e}$, and the 
normalized distribution function $g_{\rm e}\left( s,q \right)$ for the reference particle. The innocent bystander condition (iii) guarantees that no correlations are introduced between different (possibly uncorrelated or even far apart) subsets of particles. 
Property (iv), on the other hand, admits the single-particle description (\ref{eqn:CME}) not only for elementary point particles (e.g.~electrons) but also for compound objects such as atoms, molecules or even solids. 

The classicalization of the center-of-mass motion of an extended compound object of total mass $M$ can 
be approximated by the single-particle form (\ref{eqn:CME}), if 
the relative motion of the constituents around their rigidly bound equilibrium positions 
can be neglected.
The Fourier transform $\tilde{\density} \left( \vq \right) = \int \! \diff^3 x \, \density\left( \vx \right) e^{-i\vq\cdot\vx/\hbar}$
of the mass density $\density\left( \vx \right)$ of the compound modifies the rate and the phase-space distribution of the effective center-of-mass classicalization,
\begin{align}
 \frac{1}{\tau} &= \frac{1}{\tau_{\rm e}} \frac{1}{m_{\rm e}^2} \int \!\! \diff^3 s \diff^3 q \, g_{\rm e} \left( s,q \right) \left| \tilde{\density} \left( \vq \right) \right|^2, \label{eqn:tau_extended} \\
 g\left( s,\vq \right) &= \frac{\tau M^3}{\tau_{\rm e} m_{\rm e}^5} g_{\rm e} \left( \frac{M}{m_{\rm e}}s,q \right) \left| \tilde{\density} \left( \vq \right) \right|^2.
\end{align}
The effective coherence time $\tau$ depends on the relation between the size of the compound and the critical length scale $\hbar/\sigma_q$ of the reference distribution $g_{\rm e}$; 
the effective distribution $g$ 
remains normalized.
The description fails as soon as the relative motion of the constituents must be taken into account. 
The inner structure of nuclei becomes relevant 
for femtometer-scale distribution functions,
$\hbar/\sigma_q \lesssim 10\,{\rm fm} \lesssim \left( m/m_{\rm e} \right)\sigma_s $, with $m\approx 1\,$amu the nucleon mass. Here ends the domain of nonrelativistic quantum mechanics, and with it the validity of our
approach.
We thus restrict its parameters to about $\sigma_s \lesssim 20\,$pm and $\hbar/\sigma_q \gtrsim 10\,$fm,
noting that the macroscopicities will not change if these bounds are varied by a few orders of magnitude.

The restriction to a single reference distribution $g_{\rm e}(s,q)$ in (\ref{eqn:CME_N}), as opposed to individual distributions for different types of particles, yields a universal single-particle description (\ref{eqn:CME}). 
The choice of the reference mass $m_{\rm e}$ is arbitrary, since the coherence time parameter and the distribution rescale to $\tau = \tau_{\rm e} \left( m_{\rm e}/m \right)^2$ and $g\left( s,q \right) = \left( m/m_{\rm e} \right)^3 g_{\rm e} \left( m s/m_{\rm e},q \right)$ for a point particle of different mass $m$, as follows from (\ref{eqn:WN}). 
This renders the translation $\vs$ negligible for heavy objects, $m \gg m_{\rm e}$. 

In the following we will use the electron as reference particle fixing both $\tau_{\rm e}$ and $g_{\rm e}$.
Moreover, 
we take $g_{\rm e}$ to be a Gaussian distribution in $s$ and $q$,
fully specified by the standard deviations $\sigma_s$ and $\sigma_q$. 
The latter determine the main behavior of the classicalization effect;
a more involved description with additional parameters would 
complicate matters without 
significantly modifying the generic behavior.

It is remarkable that a special form of Eq.~(\ref{eqn:CME_N}), 
which we arrived at using the assumptions (i)-(iv), describes the observable consequences of the theory of ``continuous spontaneous localization'' 
(CSL) 
\cite{Ghirardi1990b,Bassi2003} if one takes $\sigma_s=0$ \cite{Vacchini2007b}. 
This shows that one can set up explicit theories which modify 
the dynamics on the level of the Schr\"odinger equation and whose observable consequences fit into the present framework \cite{Adler2007,Bassi2010,Feldmann2012,Bassi2012}. 
The stochastic Schr\"odinger equation in \cite{Ghirardi1990b,Bassi2003} 
may thus be seen as one example, but not the most general form, of a theory which yields a minimal modification in the sense described above.

\emph{Assessing superposition states.---}
The experimental demonstration of quantum coherence in a 
mechanical degree of freedom rules out a certain parameter region of the classicalizing modification, i.e.~it provides a lower bound of the 
time parameter $\tau_{\rm e}$ for any fixed value of $\sigma_s$ and $\sigma_q$.
For a superposition state in a different experiment to be more macroscopic, its demonstration must exclude a larger set of $\tau_{\rm e}$, implying that the 
modification must be even weaker. 

\begin{figure}
 \includegraphics[width=\columnwidth]{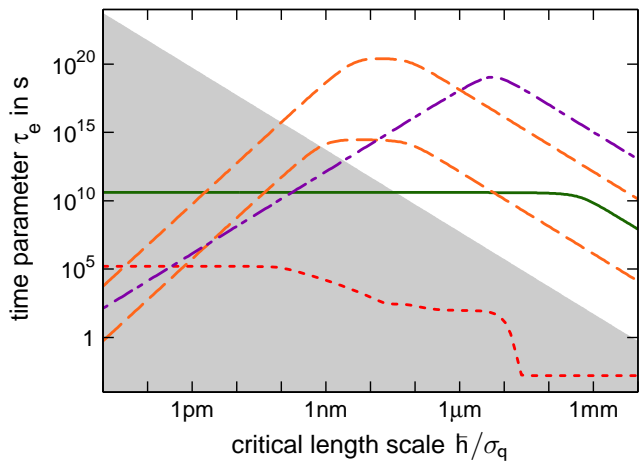}
\caption{Lower bounds on the time parameter $\tau_{\rm e}$, 
as set by various experiments. The calculations are done for the relevant range of critical length scales $\hbar/\sigma_q$, and at 
$\sigma_s=20\,$pm. 
The solid line corresponds to the atom interferometer of~\cite{Peters2001}; it 
rules out all time parameters $\tau_{\rm e}$ below the curve. Future experiments may exclude a larger set e.g.~by interference of $10^5$--$10^7\,$amu gold clusters \cite{Nimmrichter2011a_PRA_long} 
(dashed lines) or of micromirror motion \cite{Marshall2003} (dash-dotted line). The dotted line corresponds to demonstrated persistent current superpostions in a SQUID loop \cite{Friedman2000_long}. 
The shaded region represents the excluded $\tau_{\rm e}$ by a conceivable classical
measurement of less than $1\,\mu$K/s temperature increase in a Rb gas.
}
\label{fig:paramcurves}
\end{figure}

Figure~\ref{fig:paramcurves} shows the greatest excluded $\tau_{\rm e}$ for a number of different setups, as a function of the critical length scale 
$\hbar/\sigma_q$ and at fixed 
$\sigma_s=20\,$pm. The solid and the dotted curve 
correspond to exemplary modern matter-wave experiments: The interference of cesium atoms in free fall over hundreds of milliseconds (solid line)
\cite{Peters2001}, and the superposition of counterpropagating currents of $10^{14}$ superconducting electrons in a Josephson ring (dotted line)
\cite{Friedman2000_long}. The dashed and dash-dotted lines illustrate what would be achieved in proposed superposition experiments with nanoclusters \cite{Nimmrichter2011a_PRA_long} 
or micromirrors \cite{Marshall2003}. 
Whereas the value of $\sigma_s$ matters for the SQUID experiment, it is not important for the other cases due to the large masses involved. Our results then resemble the predictions of a CSL model with varying localization length $\hbar/\sigma_q$. 
Detailed results on each experiment are reported in the Appendix.

One observes a common feature of all quantum curves in Fig.~\ref{fig:paramcurves}: 
they saturate or assume a local maximum. This is because the classicalizing modification (\ref{eqn:CME_N}) is bounded in the operator norm, and any given position or momentum superposition state of a total mass $M$ thus survives at least for a time $\tau_{\rm e} \left( m_{\rm e}/M \right)^2$. 

The interferometer results (solid and dashed lines) reach the maximum where the critical length scale $\hbar/\sigma_q$ 
is comparable to the interference path separation. 
This is where the solid line saturates, in accordance with the CSL results in \cite{Feldmann2012}.
The dashed line drops at length scales smaller than the size of the interfering object, when only a fraction of its mass contributes to its center-of-mass coherence time, as given by Eq.~(\ref{eqn:tau_extended}). The latter also holds if the object is larger than the path separation (dash-dotted curve).
For smaller values of $\hbar/\sigma_q$ the slope is mainly determined by the 
mass density $\varrho \left( \vx \right)$ of the interfering object, while in the diffusive limit of large values it solely depends on $\sigma_q$. 

The superposition of persistent currents probed in the SQUID experiment \cite{Friedman2000_long} can be described by displaced Fermi spheres of Cooper-paired electrons \cite{Korsbakken2010}.
The classicalization gradually redistributes and dephases electrons between the Fermi spheres, thereby undermining the quantum coherence (see Appendix).
At large momentum spreads $\sigma_q$ the effect is governed by the 
redistribution of electrons, 
as would be the case in spontaneous localization \cite{Buffa1995}. 
The dotted curve assumes its maximum 
at a value $\sigma_q$ where the redistribution covers all electrons in the Fermi sphere. This effect vanishes for smaller $\sigma_q$ once the superconducting energy gap can no longer be overcome; 
in this limit the classicalization effect is governed by the dephasing that is induced by the position diffusion with spread $\sigma_s$.

Because of the diffusion effect inherent in 
the classicalizing modification 
there are also `classical' experiments not explicitly demonstrating quantum behavior that can be used to narrow down the range of plausible coherence time parameters.  
This is indicated by the shaded area in Fig.~\ref{fig:paramcurves} which represents the values of $\tau_{\rm e}$ excluded by an anticipated precision measurement of the temperature increase of a dilute gas of Rb atoms, 
$\sigma_q^2/2m_{\rm Rb}\tau_{\rm Rb} < 1.5 k_B\times 1\,\mu$K/s. 

\emph{Macroscopicity measure.---}
In view of the parameter bounds displayed by Fig.~\ref{fig:paramcurves} 
we suggest to 
quantify the macroscopicity of a superposition state realized in an experiment by the greatest excluded 
time parameter $\tau_{\rm e}$
of the modification (\ref{eqn:CME_N}), respecting the above parameter bounds.
To set a scale, we take the logarithm of $\tau_{\rm e}$ in units of seconds as the measure of macroscopicity,
\begin{align}
\mu = \log_{10} \left( \frac{\tau_{\rm e}}{1\,\text{s}}\right). \label{eqn:measure}
\end{align}
That is, a positive value for $\mu$ is obtained if one
demonstrates an electron to behave like a wave for more than one second, or of a proton for about a microsecond.

A simple approximate expression for the macroscopicity $\mu$ is obtained for interference experiments with point particles or with compound objects of total mass $M$, whose size is much smaller than the path separation. The single-particle modification (\ref{eqn:CME}) predicts an exponential decay of coherence with time scale $\tau = \tau_{\rm e} \left( m_{\rm e}/M \right)^2$, 
a mass dependence also obtained in the CSL case \cite{Adler2007,Bassi2010,Feldmann2012}. 
This is to be compared with the period $t$ during which coherence is maintained in the experiment. Measuring with confidence a fraction $f<1$ of the expected interference visibility, one gets
\begin{align}
 \mu = \log_{10} \left[ \left|\frac{1}{\ln f }\right| \left( \frac{M}{m_{\rm e}} \right)^2 \frac{t}{1\,{\rm s}} \right]. \label{eqn:mu_point}
\end{align}
That is to say, if one measures 30\% contrast in an interference experiment where the visibility is predicted to be, say, 60\%, then $f=0.5$ must be used in the above expression.

\begin{figure}
  \includegraphics[width=\columnwidth]{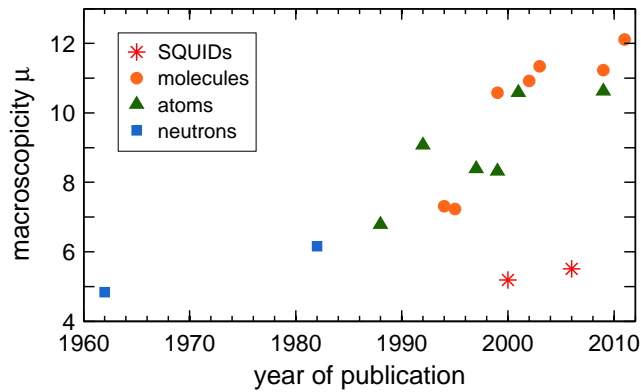}
\caption{Timeline of macroscopicities reached in quantum superposition experiments. (See the Appendix for details.) 
The squares, the triangles and the dots represent interference experiments with neutrons \cite{Maierleibnitz1962,Zeilinger1982_long}, atoms 
\cite{Keith1988,Shimizu1992,Grisenti1999,Peters2001,Chung2009_long} or atom BECs \cite{Andrews1997_short}, and molecules 
\cite{Borde1994,Chapman1995a_short,Arndt1999_short,Brezger2002_short,Hackermueller2003_short,Hornberger2009_short,Gerlich2011_short}, respectively. 
One notes that BECs do not substantially exceed the macroscopicities achieved with atom interferometers. This is due to the single-particle nature of the condensate wave function. 
The many-particle state is more involved in the case of superposition experiments with persistent supercurrent states in a large SQUID loop \cite{Friedman2000_long,Hime2006}, as represented by the stars. However, despite the large number of Cooper pairs contributing to the current superpositions in SQUIDs, such experiments lag behind in macroscopicity due to the small coherence times observed.}
\label{fig:timeline}
\end{figure} 

\emph{Macroscopicity of specific experiments.---}
In Fig.~\ref{fig:timeline} we present the macroscopicities attained in a selection of 
quantum experiments versus their publication date. They include tests of the superposition principle with neutrons, electrons, individual and Bose-condensed atoms, and molecules. Details on the calculations for specific experiments can be found in the Appendix.

\begin{table}
 \begin{center}
\begin{tabular}{l|c}
\textbf{Conceivable experiments} & $\boldsymbol\mu$\\ \hline
Oscillating micromembrane & $11.5$ \\
Hypothetical large SQUID & $14.5$ \\
Talbot-Lau interference \cite{Nimmrichter2011a_PRA_long} at $10^5\,$amu & $14.5$ \\ 
Satellite atom (Cs) interferometer \cite{Dimopoulos2009_long} & $14.5$ \\ 
Oscillating micromirror \cite{Marshall2003} & $19.0$\\
Nanosphere interference \cite{Romero-Isart2011b} & $20.5$\\
Talbot-Lau interference \cite{Nimmrichter2011a_PRA_long} at $10^8\,$amu & $23.3$ \\
Schr\"{o}dinger gedanken experiment & $\sim 57$
 \end{tabular}
 \end{center}
\caption{Expected macroscopicities for various proposed and hypothetical quantum superposition experiments.See the Appendix for details. The oscillating micromembrane setup \cite{Teufel2011} will reach the stated $\mu$-value if coherence between the zero- and one-phonon state can be observed for over 1000 oscillation cycles. For the SQUID experiment we assume a loop length of 20\,mm, a wire cross section of 100\,$\mu$m$^2$, and 1\,ms coherence time. In the gedanken experiment an idealized cat of 4\,kg is kept in a spatial superposition of 10\,cm distance for 1\,s.}
\label{tab:proposals}
\end{table} 

State-of-the-art interferometers achieve macroscopicities of up to $\mu \approx 12$, and various ideas to surpass this value with future experiments have been suggested. As can be seen from Tab.~\ref{tab:proposals}, the most promising proposals 
from the perspective of the macroscopicity measure 
employ oscillating micromirrors \cite{Marshall2003} and nanoclusters \cite{Romero-Isart2011b,Nimmrichter2011a_PRA_long}.
Their huge mass would trump a conceivable SQUID experiment with more than $10^{17}$ electrons or an atom interferometer hovering in free space with an interrogation time of one hour \cite{Dimopoulos2009_long}. 

Nevertheless, there are more than 30 orders of magnitude between experiments conceivable with present-day technology ($\mu=12$--$24$) and something as manifestly macroscopic as an ordinary house cat ($\mu \sim 57$).

\emph{Conclusion.---}
Using the measure proposed in this article 
any experiment testing the superposition principle in mechanical degrees of freedom can be quantified and compared.
By definition it answers an empirically relevant question, namely to what extent an observation serves to exclude 
minimally invasive modifications of quantum mechanics that produce classical behavior on the macroscale. As such, the measure follows directly from basic symmetry and consistency arguments, and confers physical meaning on the abstract notion of macroscopicity of a quantum system. 

The proposed measure does not depend on how a compound mechanical object is decomposed into elementary mass units.
For instance, an interfering fullerene buckyball might be described in terms of 
60 carbon atoms or equally of 1080 nucleons and electrons, and both descriptions should consistently lead to the same macroscopicity value for the overall state of the molecule.
This issue, which was not addressed in previous studies, is explicitly taken into account in our approach.
Moreover, we do not refer to specific classes of quantum states, or to preferred measurement operations or observables, rendering the measure of macroscopicity applicable to arbitrary mechanical systems.

The last 20 years have witnessed a remarkable rise in 
demonstrated macroscopicities.
Yet, new experimental strategies for quantum tests, in particular using nanoclusters and microresonators, may soon venture deeper into the macroworld.
As more and more effort is put into this field,
we may well experience an unprecedented leap towards the macroscopic domain.

\emph{Acknowledgements.---}
Support by the FWF within the projects CoQuS DK-W1210-2 and Wittgenstein Z149-N16, and by the ESF within the EuroQUASAR project MIME is gratefully acknowledged.
\pagebreak

\appendix

\onecolumngrid

\section*{APPENDIX}

Here we provide a more extensive motivation for the many-body form (5) of the modification, 
and a justification for
our estimates of the macroscopicities reached in the various experiments. We note that a more detailed analysis of the individual experimental setups might yield slightly different values, 
but would not change the overall picture.

\section{Many-particle form of the classicalizing modification}

The requirement of invariance under Galilean transformations (i) implies that the generator describing the effect of 
a minimal
modification is of the form
\begin{align}
 \cL_N \rho &= \int \!\! \diff^3 s \diff^3 v \left[ \oL \left( \vs,\vv \right) \rho \oL\da \left( \vs,\vv \right) - \frac{1}{2} \left\{ \oL\da \left( \vs,\vv \right)\oL \left( \vs,\vv \right),\rho \right\} \right] \label{eqn:ModN}.
\end{align}
The operators $ \oL \left( \vs,\vv \right)$ must satisfy 
\begin{align}
 \exp \left[-\frac{i}{\hbar} \left( \ovP \cdot\vs'-M\vv'\cdot\ovR \right) \right] \oL \left( \vs,\vv \right) \exp \left[\frac{i}{\hbar} \left( \ovP \cdot\vs'-M\vv'\cdot\ovR \right) \right] &= \exp \left[\frac{i m_{\rm e}}{\hbar} \left( \vv \cdot\vs'- \vv'\cdot\vs \right) \right] \oL \left( \vs,\vv \right) \label{eqn:LN},
\end{align}
with $M=\sum_{n=1}^N m_n$ the total mass, $m_{\rm e}$ an arbitrary reference mass, and $\ovR$, $\ovP$ the center-of-mass position and momentum operators. 
By switching to center-of-mass and relative coordinates 
it follows from Eq.~(\ref{eqn:LN}) that
the 
$ \oL \left( \vs,\vv \right)$ induce a net shift of the center-of-mass position and momentum by $m_{\rm e}\vs/M$ and $m_{\rm e}\vv$, respectively. 
However, it remains unspecified how the net shift is to be distributed amongst the $N$ constituents of the system. This freedom is constrained by the additional assumptions (ii)-(iv). 

Assumption (iii) means that the $N$-particle form (\ref{eqn:ModN}) must always reduce to the single-particle form for the $n$th particle, Eq.~(1) in the main text, if one traces over the other $N-1$ constituents, $\tr_{N-1} \left( \cL_N \rho \right) = \cL_1 \tr_{N-1} \left( \rho \right)$. That is to say, we assign to each particle species of mass $m_n$ an individual time parameter $\tau_n$ and a positive, normalized and isotropic distribution function $g_n \left( s,q \right)$. 
They constitute the free parameters of the single-particle form, as discussed in the main text. 
At the other end, assumption (iv) recovers the single-particle form (1) for the center-of-mass degree of freedom provided that the $N$ constituents are well localized at fixed equilibrium positions close to the center. We denote the corresponding classicalization parameters by $\tau^{(N)}$ and $g^{(N)}$. 
Finally, the $N$-particle  operators $ \oL \left( \vs,\vv \right)$ must be symmetrized in the case of indistinguishable particles (ii).

As a first guess one might think that the $ \oL \left( \vs,\vv \right)$ should be proportional to the unitary $N$-particle Weyl operators,
\begin{align}
 \oL \left( \vs,\vv \right) &= \sqrt{\frac{m_{\rm e}^6}{M^3 \tau^{(N)}} g^{(N)} \left( \frac{m_{\rm e}}{M}s,m_{\rm e}v  \right)} \bigotimes_{n=1}^N \exp \left[\frac{im_{\rm e}}{\hbar M} \left( m_n\vv\cdot\ovr_n - \ovp_n \cdot\vs \right) \right] \nonumber \\
 &= \sqrt{\frac{m_{\rm e}^6}{M^3 \tau^{(N)}} g^{(N)} \left( \frac{m_{\rm e}}{M}s,m_{\rm e}v  \right)} \exp \left[\frac{i}{\hbar} \left( m_{\rm e}\vv\cdot\ovR - \frac{m_{\rm e}}{M}\ovP \cdot\vs \right) \right] .
 \label{eqn:L1}
\end{align}
This way the phase-space shift would be distributed equally among all participating particles. The assumptions (ii) and (iv) would be fulfilled by construction and, due to the prefactor in (\ref{eqn:L1}), one would reobtain the center-of-mass distribution $g^{(N)}\left( s,q \right)$ in (\ref{eqn:ModN}). 
However, the operators (\ref{eqn:L1}) would leave the relative motion of any constituent subsystem entirely unaffected, irrespectively of the overall size and extension of the $N$-body system. 
Moreover, assumption (iii) is met only if $\tau^{(N)}=\tau_n$ and $g^{(N)}\left( s,q \right) = \left( m_n/M \right)^3 g_n\left( s,m_n q/M \right)$ for all $n$, so that the effective classicalization rate $1/\tau^{(N)}$ would not increase with the system size. Therefore, the operators (\ref{eqn:L1}) cannot induce classical behavior at the macroscale leaving at the same time microscopic systems unaffected. 

Rather than dividing the phase-space shift among many particles, 
one may as well compose a solution
of (\ref{eqn:LN}) from single-particle translations,
\begin{align}
 \oL \left( \vs,\vv \right) &= \sum_{n=1}^N \left( \pm \right)_{n} \sqrt{\frac{m_{\rm e}^6}{m_n^3 \tau_n} g_n \left( \frac{m_{\rm e}}{m_n}s,m_{\rm e}v  \right)} \exp \left[\frac{i}{\hbar} \left( m_{\rm e}\vv\cdot\ovr_n - \frac{m_{\rm e}}{m_n} \ovp_n \cdot\vs \right) \right].
\label{eqn:L2}
\end{align}
The sign $\left( \pm \right)_n$ may differ for distinguishable particles, and it may also depend on $s$ and $v$. These operators 
fulfill (ii) and (iii) by definition. Moreover, in the case of a 
compact compound, where $\ovr_n\approx \ovR $ and $\ovp_n \approx m_n \ovP / M$, we recover condition (iv), with the center-of-mass parameters determined by
\begin{align}
 \sqrt{\frac{m_{\rm e}^3}{M^3 \tau^{(N)}} g^{(N)} \left( \frac{m_{\rm e}}{M}s,q \right)} &= \sum_{n=1}^N \left( \pm \right)_{n} \sqrt{\frac{m_{\rm e}^3}{m_n^3 \tau_n} g_n \left( \frac{m_{\rm e}}{m_n}s,q  \right)}. \label{eqn:com_composition}
\end{align}
Each constituent contributes to the collective classicalization of the center-of-mass variables. For instance, in the case of $N$ indistinguishable particles we find that the classicalization rate is amplified by $1/\tau^{(N)} = N^2/\tau$. 
A formal proof of the operator expression (\ref{eqn:L2}) can 
be obtained in the picture of second quantization if one takes the operators $\oL\left( \vs,\vv \right)$ to be a combination of single-particle terms, i.e.~a bilinear form in the annihilation and the creation operators of a given particle species. The $N^2$-scaling of the classicalization rate then follows immediately.

In general the sign factor $\left( \pm \right)_n$ and the single-particle distributions $g_n$ might still differ from species to species. This is where the scale-invariance argument (iv) can be invoked once again: If we allowed for different signs and distribution functions for different point-particle species, we would have to define a fixed set of reference point particles, i.e.~single out a distinguished  many-body representation of any composite mechanical system. Moreover, we would end up with different descriptions of the modified time evolution of approximate point-like compound particles depending on their composition. 

In order to guarantee true scale invariance, by avoiding an ambiguous treatment of approximate point particles, one must therefore relate the extensive nature of the modification in (\ref{eqn:com_composition}) to the elementary extensive property of mechanical systems: their mass. 
We notice that the summands in (\ref{eqn:com_composition}) contribute with \emph{a priori} different mass scales $m_n$, each of which might be composed of further sub-units of mass, and so on. A unified description for any mass scale is obtained only by introducing a single reference time parameter $\tau_{\rm e}$ and distribution function $g_{\rm e}$ associated with a fixed reference mass $m_{\rm e}$. 
The composition rule (\ref{eqn:com_composition}) then holds naturally once we identify 
\begin{align}
 \left( \pm \right)_{n} \sqrt{\frac{m_{\rm e}^3}{m_n^3 \tau_n} g_n \left( \frac{m_{\rm e}}{m_n}s,q  \right)} &= \frac{m_n}{m_{\rm e}} \sqrt{\frac{1}{\tau_{\rm e}} g_{\rm e} \left( s,q \right) }\,.
\end{align}
The classicalization effect now scales uniformly with mass, irrespectively of the types of particles involved.

\section{Diffraction at gratings and double-slits}

The macroscopicity observed in the matter-wave diffraction experiments at gratings and double-slits \cite{Maierleibnitz1962,Zeilinger1982_long,Keith1988,Shimizu1992a,Grisenti1999,Arndt1999_short} can be estimated from the height of the first order diffraction peak in the recorded signals.
Given the transmission function $t(x)$ of a one-dimensional $N$-slit-grating with slit distance $d$ and opening width $w$, we find the interference signal of a monochromatic point source (velocity $v_z$) in the paraxial approximation
\begin{align}
 I_{v_z} (x) &\propto \int \!\! \diff x_1 \diff x_2 \, R\left( x_1-x_2 \right) t(x_1) t^{*} (x_2) \, \exp \left\{ \frac{im}{2\hbar} \left[ \frac{x_1^2-x_2^2}{T} - \frac{2x(x_1-x_2)}{T_2} \right] \right\}. \label{eqn:diffraction_ideal}
\end{align}
The time $T=T_1 T_2 / (T_1+T_2)$ is determined by the times-of-flight $T_1$ from the source to the grating and $T_2$ from the grating to the screen. In a horizontal alignment they are related to the respective distances $L_{1,2}=v_z T_{1,2}$. The signal (\ref{eqn:diffraction_ideal}) must be averaged with respect to the distributions of the velocities $v_z$, and over the extensions $S$ and $D$ of the source slit and the detector. The single-particle classicalization as described by equation (1) in the main text
blurs the interference signal by the factor
\begin{align}
R(x) &= \exp \left\{ \frac{T_1}{\tau} \int_0^{1} \!\! \diff z \left[ \widetilde{g}_{\rm 1D} \left( x z, \frac{mx}{T_1}  \right) -1 \right] + \frac{T_2}{\tau} \int_0^{1} \!\! \diff z \left[ \widetilde{g}_{\rm 1D} \left( x z, \frac{mx}{T_2}  \right)-1 \right] \right\},
\end{align}
involving the reduced 
Fourier transform $\widetilde{g}_{\rm 1D} (x,p) = \widetilde{g} \left( x \ex, p \ex \right)$ of the distribution function $g$. 
The unperturbed fringe pattern exhibits diffraction maxima at screen coordinates close to integer multiples of $x=h T_2 / m d$. 
The classicalization affects them strongest if the contributing interference paths are completely resolved
by the critical length scale $\hbar/\sigma_q$.
In this limit a flat background is added to the overall signal, and
the longitudinal velocity distribution affects only weakly 
the reduction of 
the first diffraction maximum; the latter can thus be used to extract the macroscopicity $\mu$ for each experiment according to equation (9) of the main text. 
The required parameter $f$ is estimated by the ratio of the measured height of the diffraction peak and its unperturbed theoretical value, both normalized to the integrated signal. 
The following table contains all required parameters of the different experiments, including references to the data used for the comparison.

\small
\begin{center}
 \begin{tabular}{cc|cccccccccc|cc}
  \textbf{Ref.} & Fig. & $L_1$/m\, & $L_2$/m\, & $N$\, & $d$/$\mu$m\, & $w/d$\, & $S$/$\mu$m\, & $D$/$\mu$m\, & $\la v_z\ra$/$\tfrac{\rm m}{\rm s}$\, & $\Delta v_z / \la v_z\ra$\, & $m$/amu\, & $f$\,  & $\mu$\, \\ \hline \vspace*{-3mm} & & & & & & & & & & & & & \\
 \cite{Maierleibnitz1962} & 9 & 4.0 & 5.7 & 2 & 107 & -- & 10 & 30 & 907 & -- & 1 & 0.6 & 4.8 \\
\cite{Zeilinger1982_long} & 7 & 5.0 & 5.0 & 2 & 126 & 0.17 & 15 & 30 & 216 & 0.05 & 1 & 0.9 & 6.2 \\
\cite{Keith1988} & 2b & 1.0 & 1.5 & 50 & 0.2 & 0.5 & 10 & 25 & 1000 & 0.12 & 23 & 0.5 & 6.8 \\
\cite{Shimizu1992a} & 3b & 0.08 & 0.11 & 2 & 6 & 0.33 & 20 & 20 & -- & -- & 20 & 0.8 & 9.1 \\
\cite{Grisenti1999} & 1 & 0.45 & 0.52 & 100 & 0.1 & 0.43 & 10 & 25 & 396 & 0.1 & 84 & 0.8 & 8.3 \\
\cite{Arndt1999_short} & 2a & 1.14 & 1.25 & 100 & 0.1 & 0.38 & 10 & 8 & 226 & 0.6 & 720 & 0.6 & 10.6 \\
 \end{tabular}
\end{center}

\normalsize
The neutron interference at a biprism observed in \cite{Maierleibnitz1962} can be related to the coherent superposition of two virtual sources separated by $d=107\,\mu$m, 9.7\,m away from the detector. The resulting fringe pattern thus resembles a double-slit pattern. The authors of \cite{Maierleibnitz1962}  present the measured data and a fitted theory curve in Fig.~9. The data deviates from the predicted height of 200\,a.u.~of the first diffraction order by roughly 50\,a.u. Subtracting a dark count rate of 60\,a.u.~leads to the estimate $f \sim 1 - 50/140 \approx 0.6$.
The experiment \cite{Shimizu1992a} is a vertically aligned interferometer, where neon atoms are released from a trap. They fall through a double-slit and into a detector within $t\approx 200\,$ms. We obtain $f$ by comparing the measured diffraction peak in Fig.~3b with the theoretical model of \cite{Shimizu1992a} in Fig.~3g. 
For \cite{Maierleibnitz1962,Zeilinger1982_long,Keith1988,Shimizu1992a,Grisenti1999} we use a Gaussian velocity distribution with the mean $\la v_z \ra$ and the FWHM $\Delta v_z$, as specified in the table. For \cite{Arndt1999_short} we use the distribution provided in the article, and we account for the special detection scheme by replacing the detector slit by a Gaussian laser focus of waist $D$. The dispersive interaction between the particles and the grating walls is taken into account for \cite{Grisenti1999,Arndt1999_short} by reducing the effective slit opening size. 

The macroscopicity of the proposed optical double-slit experiment \cite{Romero-Isart2011b} with silica nanospheres is estimated by considering the Fourier amplitude which corresponds to the expected double-slit fringe oscillation (\ref{eqn:diffraction_ideal}). 
The classicalization modifies it by the factor $R\left( d \right)$, with the largest proposed value for the slit distance $d=52\,$nm. We evaluate the macroscopicity by modeling the particles as homogeneous spheres ($\varrho = 2200\,$kg/m$^3$) of $20\,$nm radius (see equations (6) and (7) of the main text),
and by assuming that at least 50\% of the fringe amplitude is observed.

Talbot-Lau interference with molecules and clusters \cite{Brezger2002_short,Hackermueller2003_short,Hornberger2009_short,Gerlich2011_short,Nimmrichter2011a_PRA_long} can be treated in a similar manner. 
The sinusoidal fringe visibility $\V_{\sin}$ of a symmetric setup ($T_1=T_2=T$) is reduced to $R\left( hT/md \right) \V_{\sin}$. 
Judging from the error bars at high visibilities, we assume that the measurements are compatible with at least 90\% of the prediction for C$_{70}$ molecules (Fig.~3 in \cite{Brezger2002_short}), 90\% for C$_{60}$F$_{48}$ (Fig.~5 in \cite{Hornberger2009_short}), and 80\% for PFNS8 (Fig.~4b in \cite{Gerlich2011_short}). 
The fringe pattern observed in \cite{Hackermueller2003_short} with C$_{60}$F$_{48}$ (Fig.~4) corresponds to 75\% of the predicted visibility.

\section{Ramsey-Bord\'{e} interference with I$_2$ molecules}

In the experiment \cite{Borde1994} a beam of I$_2$ molecules ($m=254\,$amu) passes two pairs of counterpropagating running-wave laser beams, 
as described in detail in \cite{Borde1984}. 
Two paths through the setup contribute to the recorded Ramsey fringe pattern, as shown in Fig.~2 of Ref.~\cite{Borde1994}. 
We must include, however, a significant contribution from two further paths to the signal \cite{Borde1984}. 
Their interference is washed out over the transverse velocity distribution of the molecule beam. 
We assume that they add an offset to the most pronounced central fringe of the (*)-curve in Fig.~2 of \cite{Borde1994}. If all four paths contribute by roughly the same weight we must halve the offset of the central fringe in the diagram, which yields a two-path fringe visibility of $f \approx 400\,{\rm a.u.}/(2400-1200)\, {\rm a.u.} = 0.33$. The passage time is determined by the $(35+2)\,$mm length of the interferometer and the mean molecular velocity of $350\,$m/s. This yields $\mu = 7.3$, according to equation (9) of the main text.

\section{Mach-Zehnder-type interference}

The two atom interferometers featuring the greatest macroscopicity in Fig.~2 \cite{Peters2001,Chung2009_long}, as well as the proposed satellite atom interferometer \cite{Dimopoulos2009_long} listed in Tab.~1, 
are optical Mach-Zehnder-type geometries, which could in principle yield close to 100\% fringe contrast. We use the recorded fringe visibilities $f=0.62$ (Fig.~19 in \cite{Peters2001}) and $f=0.33$ (Fig.~3 in \cite{Chung2009_long}), and a hypothetical value of $f=0.5$ for the proposal \cite{Dimopoulos2009_long}. In all three cases the interfering particles are $^{133}$Cs atoms, and the interrogation time is given by twice the pulse separation time $T$. The respective values are $T=160\,$ms, 400\,ms, and 2000\,s.

In the Na$_2$ molecule interferometer \cite{Chapman1995a_short} the Mach-Zehnder geometry is realized with three material diffraction gratings. The total length of the interferometer including beam collimation is about $2.1\,$m \cite{Chapman1995_short,Berman1997}, and the molecules pass it at a mean velocity of $820\,$m/s. This yields an interrogation time of $2.6\,$ms. 
The maximally possible contrast is limited by two factors: First, the different weights of the interference paths, which correspond to the zeroth and the first 
diffraction order at the first grating; they are given by $P_1/P_0 = \sinc \left( 0.3 \pi \right)/1 = 0.74$. Second, the modulation of the interference pattern by the third grating mask; it contributes a factor of $\sinc\left( 0.3 \pi \right)=0.86$ to the fringe amplitude, assuming a sinusoidal fringe pattern and a grating opening fraction of 30\%. The detected contrast is thus limited to below 85\%. 
We extract a measured fringe contrast of about 30\% from the inset of Fig.~4 in \cite{Chapman1995a_short}, i.e.~$f=0.35$, which leads to $\mu = 7.2$.

\section{Oscillating microresonators}

The authors of \cite{Marshall2003} propose to create a quantum superposition state of an oscillating micromirror by entangling it with a single cavity photon in one arm of a Michelson interferometer. If coherence is maintained in the mirror motion during one period of oscillation photon interference fringes should be observed at 100\% contrast.

The classicalization master equation
(equation (1) in the main text) 
for the harmonic mirror motion can be integrated explicitly. We find that it reduces the fringe visibility after one oscillation period $2\pi/\omega_m$ by the factor
\begin{align} 
 R &= \exp \left\{ \int_0^{2\pi} \!\! \frac{\diff \xi}{\omega_m \tau} \left[ \widetilde{g}_{\rm 1D} \left( 2 \kappa x_0 \sin^2 \frac{\xi}{2},\frac{2\hbar\kappa}{x_0} \sin \xi \right) -1 \right] \right\}
\end{align}
with $\widetilde{g}_{\rm 1D}$ the reduced Fourier transform of the distribution $g$. The latter and the time parameter $\tau$ are given by equations (6) and (7) in the main text. 
The micromirror is modeled as a homogeneous cube of mass density $\varrho=2300\,$kg/m$^3$, $b=10\,\mu$m edge length and a mass of $M=\varrho b^3 = 2.3\,$ng, which yields
\begin{align} 
 \frac{1}{\tau} &= \frac{1}{\tau_e} \left( \frac{M}{m_e} \right)^2 \gamma^3, \\
 g_{\rm 1D} \left( s,q \right) &= \gamma^{-1} \frac{M}{2\pi m_e \sigma_s \sigma_q} \exp \left( - \frac{M^2 s^2}{2 m_e^2 \sigma_s^2} - \frac{q^2}{2\sigma_q^2} \right) \sinc^2 \left( \frac{qb}{2\hbar} \right) \label{eqn:g1D_canti},
\end{align}
with the $\sigma_q$-dependent geometry factor
\begin{align}
 \gamma &= 2 \left( \frac{\sigma_q b}{\hbar} \right)^{-2} \left[ \exp \left( - \frac{\sigma_q^2 b^2}{2\hbar^2} \right) + \sqrt{\frac{\pi}{2}} \frac{\sigma_q b}{\hbar} \erf \left( \frac{\sigma_q b}{\sqrt{2} \hbar} \right) - 1  \right].
\end{align}
The authors presume a frequency $\omega_m / 2\pi = 500\,$Hz, a ground state oscillation amplitude of $x_0=170\,$fm and a photon-mirror coupling strength of $\kappa = 1.63$. We find $\mu = 19.0$ for a measured 50\% fidelity.

For the hypothetical superposition experiment with an oscillating Al micromembrane, as listed in Table~I of the main text, we use the parameters given in \cite{Teufel2011}. The membrane mass $M=48\,$pg and the mechanical frequency $\omega_m/2\pi = 10.56\,$MHz yield a tiny ground state amplitude of $x_0 = \sqrt{2\hbar/M \omega_m} = 8\,$fm. To give a good upper estimate of the macroscopicity of such an experiment we thus approximate the flexural mode of the membrane by a axial center-of-mass vibration of a homogeneous disc of thickness $b=100\,$nm and radius $R=7.5\,\mu$m. We obtain the effective distribution $g_{\rm 1D} \left( s,q \right)$ of (\ref{eqn:g1D_canti}) and a time parameter
\begin{align}
 \frac{1}{\tau} &= \frac{2 \gamma}{\tau_e} \left( \frac{M}{m_e} \right)^2 \left( \frac{\sigma_q R}{\hbar} \right)^{-2} \exp \left( - \frac{\sigma_q^2 R^2}{\hbar^2} \right) \left[ \exp \left( \frac{\sigma_q^2 R^2}{\hbar^2} \right) - I_0 \left( \frac{\sigma_q^2 R^2}{\hbar^2} \right) - I_1 \left( \frac{\sigma_q^2 R^2}{\hbar^2} \right) \right],
\end{align}
with $I_{0,1}$ the modified Bessel functions.

We assume that the membrane is prepared in the superposition state $|\psi\ra = \left( |0\ra + |1\ra \right)/\sqrt{2}$ of the zero- and the one-phonon eigenstate, and that the associated nondiagonal matrix element $\la 1|\rho_t|0\ra$ does not decay by more than 50\% after a time $t=2\pi n/\omega_m$ which corresponds to $n=1000$ oscillation cycles. Due to the large mass $M$ and the small amplitude $x_0$ we may neglect the position spread $\sigma_s$ in (\ref{eqn:g1D_canti}), and we may Taylor-expand its Fourier transform $\widetilde{g}_{\rm 1D} \left( x,p \right)$ to lowest order in $x$. 
With this we arrive at the explicit condition
\begin{align}
 \frac{\la 1 |\rho_t |0\ra}{\la 1|\psi \ra \la \psi|0\ra} &= \left( \frac{2\pi n}{\omega_m \tau} \frac{x_0^2}{\gamma b^2}\left[ 1 - \exp \left( - \frac{\sigma_q^2 b^2}{2\hbar^2} \right)\right] + 1 \right)^{-2} \geq 50\%,
\end{align}
which leads to a macroscopicity of $\mu = 11.5$.

\section{BEC interference}

The interference of two sodium BECs observed in \cite{Andrews1997_short} is modeled using a second quantization phase-space picture in \cite{Wallis1997_long}. Following the same line we define a second quantization form of the characteristic function,
\begin{align}
 \hat{\chi} \left( x,p \right) &= \int \!\! \diff x_0\, e^{ipx_0/\hbar} \hat{\psi}\da \left( x_0 + \frac{x}{2} \right) \hat{\psi} \left( x_0 - \frac{x}{2} \right). \label{eqn:charfunc_BEC}
\end{align}
De Broglie interference of trapped BECs is observed as a fringe pattern in the time-evolved single-particle density $\hat{n} \left( x \right) = \hat{\psi}\da \left( x \right) \hat{\psi} \left( x \right)$ for each individual run of the experiment. The fringe visibility is given by the corresponding Fourier component of $\hat{n} \left( x \right)$,
\begin{align}
 \hat{\chi} \left( 0, \frac{h}{\lambda} \right) &= \int \!\! \diff x_0\, e^{2\pi i x_0/\lambda} \hat{n} \left( x_0 \right),
\end{align}
where $\lambda$ denotes the fringe spacing.
The pattern observed in each run of the experiment \cite{Andrews1997_short} can be assessed in the case of non-interacting bosons by replacing the annihilation operator $\hat{\psi}\left( x \right)$ with the collective wave function $\psi \left( x \right)$ of the two trapped condensates. 
A free evolution of (\ref{eqn:charfunc_BEC}) by the time $t$ then yields the visibility $\chi \left( 0, h/\lambda \right)$ of the resulting interference pattern.  The fringe spacing $\lambda=$ should be modified to account for interactions in the BEC \cite{Wallis1997_long}.

The second quantization form of the reduced $N$-particle operators (4) reads as
\begin{align}
 \oW \left( s,q \right) &= \frac{m}{m_e} \int \!\! \diff x \, e^{-iqx/\hbar} \hat{\psi}\da \left( x \right) \hat{\psi} \left( x + \frac{m_e}{m}s \right).
\end{align}
A straightforward calculation reveals that (\ref{eqn:charfunc_BEC}) then classicalizes at the rate
\begin{align}
 \cL \, \hat{\chi} \left( x,p \right) &= - \left( \frac{m}{m_e} \right)^2 \frac{1}{\tau_e} \left[ 1 - \frac{m}{m_e} \int \!\! \diff s \, \diff q \,\, g_e \left( \frac{m}{m_e}s,q \right) e^{i(qx-ps)/\hbar} \right]\hat{\chi} \left( x,p \right) \label{eqn:BEC_charfunc_class}
\end{align}
of a single atom. We therefore estimate the macroscopicity from equation (9) of the main text. The authors of \cite{Andrews1997_short} observed 
about $f=75$\% interference contrast in a sodium BEC after a time-of-flight of $t=40\,$ms; this yields $\mu=8.4$. 
In the experiment \cite{Jo2007} the phase sensitivity of the interferometer was increased, but at an interference contrast of only 15\% after 200\,ms, which leads to $\mu=8.3$ (not discussed in the main text).

Modern-day experiments with multi-component BECs make use of nonlinear interactions and number squeezing to increase the coherence time and the phase sensitivity employing internal atomic states \cite{Riedel2010,Gross2010,Egorov2011}. 
The use of such techniques in interference experiments with spatially split BECs would only yield a macroscopicity $\mu$ that is comparable to single-atom interferometers. 
This is due to the fact that the single-particle nature of the classicalizing effect (\ref{eqn:BEC_charfunc_class}) holds irrespectively of whether nonlinear interactions modify the coherent time evolution of the condensate wave function. Larger values of $\mu$ could be achieved by increasing the fringe visibility and the time-of-flight in both single-atom and BEC experiments, possibly carried out in a microgravity environment.

\section{SQUID interference}

For the case of SQUID experiments we obtain the exclusion curve of the classicalization parameters in Fig.~1 and the $\mu$-values in Fig.~2 and Tab.~1 
by estimating the decay rate of a superposition state of macroscopically different supercurrents, i.e.~different phases across the 
junctions in a Josephson loop. 
This was studied theoretically for spontaneous localization models in \cite{Buffa1995}, whose observable consequences are a special case of the classicalizing modification discussed here \cite{Vacchini2007b}.

A state of finite current density $|\vj\ra$ in a solid with electron density $n_e$ is described by a Fermi sphere, displaced by the momentum $\hbar \vk_{\vj}$, $\vj = n_e e \hbar \vk_{\vj} / m_e$. The undisplaced state $|0\ra$ is taken to be the BCS ground state of the superconductor \cite{Bardeen1957,Leggett2006SC}. It is characterized by the probability amplitudes $v_{\vk}$ ($u_{\vk}=\sqrt{1-v_{\vk}^2}$) of a Cooper pair $(\vk \uparrow, -\vk \downarrow)$ being occupied (unoccupied),
\begin{align}
 v_{\vk} &= \frac{1}{2} \left( 1 - \frac{k^2 - k_F^2}{\sqrt{ \left( k^2 - k_F^2 \right)^2 + \left( 2 m_e \Delta_{\vk}/\hbar^2 \right)^2 }} \right).
\end{align}
Here, $\hbar k_F = m_e v_F$ denotes the Fermi momentum and $\Delta_{\vk}$ the pairing energy. The latter is approximated in the usual way by the zero-temperature energy gap $\Delta=1.76k_B T_c$ for electrons close to the Fermi level, $|k^2-k_F^2| \leq 2m_e \omega_D / \hbar$, and zero otherwise. 
The term $\omega_D$ denotes the Debye cutoff frequency of the material. We use the literature values $k_F = 1.18\,$\AA$^{-1}$, $\Delta=1.44\,$meV, $\hbar\omega_D = 23.7\,$meV for Nb, and $k_F = 1.74\,$\AA$^{-1}$, $\Delta=0.17\,$meV, $\hbar\omega_D = 36.9\,$meV for Al, respectively \cite{Kittel1996Solid,Korsbakken2010}.

The second quantization form of the classicalization operators (equation (4) in the main text) for electrons reads as
\begin{align}
 \oW_e \left( \vs, \hbar \vq \right) &= \sum_{\sigma=\uparrow,\downarrow} \sum_{\vk} e^{i\vk \cdot \vs} \oa\da_\sigma \left( \vk \right) \oa_\sigma \left( \vk + \vq \right).
\end{align}
The sum covers all discrete electron momentum states in a given volume $V$,  each state occupying the elementary cell $(2\pi\hbar)^3/V$ in momentum space. The classicalization kick distribution $g_e \left( s,q \right)$ must be discretized accordingly. We find that a superposition state $\rho$ of two distinct supercurrents $\vj_1$ and $\vj_2$ decays at a rate
\begin{align}
 \Gamma &\approx \left. - \frac{\partial_t \la \vj_1 |\rho(t)|\vj_2 \ra}{\la \vj_1 |\rho(t)|\vj_2 \ra} \right|_{t=0} = \Gamma_{\rm diff} + \Gamma_{\rm deph}
\end{align}
due to classicalization. This assumes that $\la \vj_1 |\vj_2 \ra = 0$, and that the net number $\delta N$ of electrons occupying different states in each superposition branch \cite{Korsbakken2010} is large. The decay rate splits into two contributions. The first one is related to momentum diffusion, which requires that at least one elementary unit of momentum $2\pi\hbar / V^{1/3}$ is transferred. In the continuum limit $\sum_{\vk} \to V/(2\pi)^3 \int \! \diff^3 k$ we find
\begin{align}
\Gamma_{\rm diff} &= \frac{2 V \hbar^3}{(2\pi)^3 \tau_e} \!\! \iiint\limits_{q> \pi/V^{1/3}} \!\!\! \diff^3 s \, \diff^3 q \,\diff^3 k \,\, g_e \left( s, \hbar q \right) \, u_{\vk} v_{\vk + \vq} \left( u_{\vk} v_{\vk + \vq} + v_{\vk} u_{\vk + \vq} e^{i (2\vk + \vq)\cdot \vs} \right).
\end{align}
The expression is ultimately bounded by $\Gamma_{\rm diff} \leq N/\tau_e$ in the limit of arbitrarily strong momentum kicks, when all $N=n_e V$ conducting electrons can be transferred from one branch of the superposition to the other. It does not depend on the actual value of the supercurrents.

The second contribution represents the dephasing that comes from the classicalization-induced position kicks, when no momentum redistribution of the electrons takes place,
\begin{align}
\Gamma_{\rm deph} &= \frac{4 V^2 \hbar^3}{(2\pi)^6 \tau_e} \!\!\! \iint\limits_{q \leq \pi/V^{1/3}} \!\!\!\!\! \diff^3 s \, \diff^3 q \,\, g_e \left( s, \hbar q \right)  \left( 1 - e^{i\delta \vk \cdot \vs} \right) \left| \int \!\! \diff^3 k \,\, v_{\vk}^2 \, e^{i\vk\cdot\vs} \right|^2.
\end{align}
Here, $\hbar \delta \vk = m_e (\vj_1 - \vj_2) / n_e e$ denotes the difference in momentum displacement of the two current branches. It is orders of magnitude smaller than the Fermi momentum, and $|\delta \vk \cdot \vs| \ll 1$ holds for any reasonable kick distribution $g_e$. Hence the dephasing contribution scales quadratically with the net difference in occupation of the two displaced Fermi spheres, $\delta N = 4 N |\delta \vk| / 3 k_F$. 
While this may potentially be significant for large SQUID geometries, the diffusion contribution dominates in all existing real-size experiments.

Experimentally measured coherence times $T_2$ of such current superpositions provide an upper bound for the decay rate $\Gamma$. We estimate $T_2$ by the smallest observed frequency splitting in the experiments \cite{Friedman2000_long} ($T_2\approx 1\,$ns) and \cite{Hime2006} ($T_2\approx 10\,$ns); the authors of \cite{Wal2000_short} estimate $T_2 \approx 15\,$ns.
Classicalization parameters which lead to $\Gamma > 1/T_2$ are then excluded by each experiment. This yields the SQUID curve in Fig.~1, as well as the $\mu$-values plotted in Fig.~2; 
the latter are computed with the boundary condition $\sigma_s \leq 1\,\text{ \AA} \leq \hbar / \sigma_q $, as discussed in the main text. 
The superconducting loop is spanned by $L=560\,\mu$m of Nb in \cite{Friedman2000_long}, $20\,\mu$m of Al in \cite{Wal2000_short}, and $180\,\mu$m of Al in \cite{Hime2006}. We assume the respective material cross sections as $5\,\mu$m$^2$, $36000\,$nm$^2$, and $1\,\mu$m$^2$. The experiment \cite{Wal2000_short} yields a smaller macroscopicity, $\mu = 3.3$, than \cite{Friedman2000_long} ($\mu=5.2$) due to its smaller ring geometry. Only the greater value is included in Fig.~2. 
The large hypothetical SQUID in Tab.~1 of the main text is a 20\,mm loop of $100\,\mu$m$^2$ cross section with a coherence time of 1\,ms.
 
The actual values of the supercurrents do not influence the $\mu$-values, since the dephasing contribution is negligible in all cases. 
We use a current difference of $I_1-I_2 = 3\,\mu$A for Fig.~1, 
as given in \cite{Korsbakken2010}.

\section{Schr\"{o}dinger's gedankenexperiment}

In our version of the famous gedankenexperiment, as listed in Tab.~1 of the main text, we consider the hypothetical superposition state of an ideal cat sitting at two places $\vx_1$ and $\vx_2$ that are 10\,cm apart. The center-of-mass coherence of the cat then decays like
\begin{align}
 \frac{\partial_t \la \vx_1|\rho|\vx_2\ra}{\la \vx_1|\rho|\vx_2\ra} &= \frac{1}{\tau} \int \!\! \diff^3 s \, \diff^3 q \,\, g\left( s,\vq \right) \left( e^{i\vq\cdot (\vx_2 - \vx_1)/\hbar} -1 \right)
\end{align}
due to classicalization. We have neglected the weak position diffusion in the classicalization master equation~(1) here. The mean coherence time of this state shall be 1\,s. In order to evaluate the above decay rate using the $\tau$ and the $g$ of a compound, as defined by (6) and (7) in the main text,
we model the cat as a homogeneous sphere of water with a mass of 4\,kg.

\twocolumngrid


%

\end{document}